\documentclass{emulateapj}

\usepackage{graphics,psfig}

\newcommand{\msun}{$M_{\sun}$\,}

\shorttitle{Spiral arms in M\,82}
\shortauthors{Mayya et al.}

\begin{document}

\title{The discovery of spiral arms in the starburst galaxy M\,82}
\author{Y. D. Mayya, L. Carrasco,  and  A. Luna}
\affil{Instituto Nacional de Astrof\'{\i}sica, Optica y Electr\'onica, 
Luis Enrique Erro 1, Tonantzintla, C.P. 72840, Puebla, Mexico}
\email{ydm@inaoep.mx, carrasco@inaoep.mx, aluna@inaoep.mx}

\begin{abstract}
We report the discovery of two symmetric spiral arms in the 
near-infrared (NIR) images of the starburst galaxy M\,82. The spiral arms are
recovered when an axi-symmetric exponential disk is subtracted from the NIR 
images. The arms emerge from the ends of the NIR bar and can be traced up to 
three disk scalelengths. The winding of the arms is consistent with an $m=2$ 
logarithmic spiral mode of pitch angle $14^\circ$.
The arms are bluer than the disk in spite of their detection on the NIR images.
If the northern side of the galaxy is nearer to us, as is normally assumed, 
the observed sense of rotation implies trailing arms.
The nearly edge-on orientation, high disk surface brightness, and the 
presence of a complex network of dusty filaments in the optical images, are 
responsible for the lack of detection of the arms in previous studies.
\end{abstract}


\section{Introduction}

M\,82 is a nearby edge-on galaxy (D = 3.63 Mpc, image scale 
17.6~pc\,arcsec$^{-1}$; Freedman et al. 1994), which has revealed interesting 
phenomena at every spatial scale investigated; on the largest scale, a bridge 
of intergalactic gas, spanning over 20~kpc, connects M\,82 with its 
neighbor M\,81 \citep{Gott77}. 
The external part of the stellar disk (radius$\sim$5~kpc) is warped in 
the direction of the HI streamers detected by \citet{Yun93}. The central 
500~pc of the galaxy harbors a starburst, the prototype 
for the starburst phenomenon \citep{Riek80}. Starburst-driven galactic 
superwinds were detected upto distances of 10~kpc above the plane of the 
galaxy \citep{Lehn99}. Surrounding the active starburst region, there are
a molecular ring of 400~pc radius \citep{Shen95}, and a 
near-infrared (NIR) bar of $\sim$1~kpc length \citep{Tele91}. A 
network of dusty filaments dominate the optical appearance of the galaxy. 
Morphologically, M\,82 belongs to the Irr II class of galaxies \citep{Holm50},
later called I0 by \citet{deVa59} and ``Amorphous'' by \citet{Sand79}.
M\,82 is a low mass galaxy ($\sim10^{10}$\,\msun), with most of its mass 
concentrated within the central 2~kpc \citep{Sofu98}. 

There is overwhelming evidence that M\,82 has recently undergone close 
encounters with its massive neighbor M\,81 and the dwarf galaxy NGC\,3077 
\citep{Yun94}. Detailed N-body simulations of interacting systems have shown 
that the onset of gaseous infall is intimately related to the formation of 
global bars and/or spiral arms (i.e. $m=2$ mode), which help drive the
gas towards the central regions \citep{Toom72,Barn92,Miho99}. The presence 
of the NIR bar in M\,82 is in agreement with this general picture. However, in
M\,82 the bar occupies only one tenth of the optical disk, suggesting that 
there may be hidden components such as spiral arms extending over the 
entire disk. Given the large amount of extinction in the optical bands,  
a careful analysis of NIR images is needed to uncover such features.
\citet{Ichi95} carried out NIR surface photometry of M\,82, and 
found distortions in the isophotes outside the central bar, which they assumed 
to be related to the warp of the external disk. In this {\it Letter}, using 
new deep NIR images, we comprehensively show that these isophotal distortions 
in M\,82 are due to the presence of a two-armed logarithmic spiral structure.

In Section 2, we describe the observations and the techniques adopted in their
analysis. In Section 3 and 4, we characterize the properties of the disk and 
spiral arms. Finally, in Section 5, we discuss the morphological type of M\,82.

\section{New NIR Observations and analysis}

New NIR imaging of M\,82 in the $J,H$ and $K$ bands were obtained during
2004 March 30 and 31, with the CAnanea Near Infrared CAmera (CANICA) 
available at the 2.1-m telescope of the Observatorio Astrof\'{\i}sico 
Guillermo Haro in Cananea, Sonora. CANICA hosts a $1024\times1024$ format, 
HAWAII array. The image scale and field of view of the camera are 
0.32\arcsec\,pixel$^{-1}$ and $\sim5.5$\arcmin$\times$5.5\arcmin, respectively.
A sequence of object and sky images were taken, with the final co-added
images corresponding to integration times on the object of 18, 6 and 12.5 
minutes in $J,H$ and $K$ bands, respectively. The FWHM of the seeing disk
was $\sim$2.5\arcsec. The galaxy slightly overfills the detector array, 
and the outermost isophotes in our images correspond to surface brightness 
levels of $19.5\pm0.1$, $19.0\pm0.3$ and $18.5\pm0.2$\,mag\,arcsec$^{-2}$ in 
the $J,H$ and $K$ bands, respectively. The signal-to-noise ratio (SNR) 
is in excess of 50 for the azimuthally averaged intensities. Photometric 
standard stars from \citet{Hunt98} were observed to calibrate the images. The 
resulting magnitudes for the central 35\arcsec\ aperture agree within 
0.05~mag with those extracted over the same aperture from the 2MASS archival 
images. Further information on the CANICA instrumentation and the reduction
package can be found in \citet{Carr05}.
We complemented our NIR data with $B$-band images \citep{Marc01},
archived at the NASA Extragalactic Database (NED). 

\section{Surface photometric analysis}

The optical appearance of M\,82 is dominated by bright star-forming knots 
interspersed by the dusty filaments. However, in the NIR, the light 
distribution is smooth, rendering the images ideal for studying the properties 
of the stellar disk. \citet{Ichi95} analyzed NIR images, and found variations 
of the position angle (P.A.) of the semi-major axis and axial ratio $b/a$, 
as a function of radius. In cases like this, the determination of the 
inclination of the disk to the line of sight is highly uncertain. Hence
\citet{Pier88} have suggested the use of a characteristic $b/a$ value jointly
with an intrinsic ratio for the scaleheight to scalelength of 0.20. 
The range of the observed $b/a$ ratios provides an estimator of error for the 
inclination angle. For M\,82, we determine a characteristic $b/a=0.30$, 
corresponding to an inclination angle of $77\pm3^\circ$.
The surface photometry was carried out with the ELLIPSE task in IRAF/STSDAS. 

We obtained one-dimensional (1-d) intensity profiles in $J,H$ and $K$ bands by 
azimuthally averaging the intensities in successive annular regions. The 1-d 
intensity profiles in the outer parts of the galaxy follow an exponential law 
as expected for a stellar disk. A fit to the $K$-band profile outside a radius 
of 60\arcsec\ yields a scalelength of $47\pm3$\arcsec ($825\pm50$~pc), in good 
agreement with the values obtained by \citet{Ichi95}. The error on the 
scalelength was estimated by fitting an exponential function to different 1-d 
profiles, each profile obtained by varying $b/a$ between 0.28--0.32, and P.A. 
between 62--65$^\circ$. The scalelengths in the $J$ and $H$-bands are identical
to that in the $K$-band within the quoted errors. 
\begin{figure}
\epsscale{1.15}
\plotone{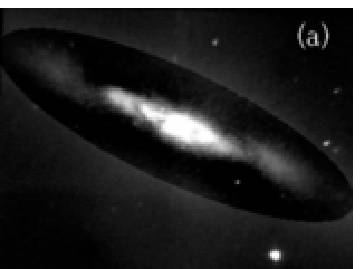}
\plotone{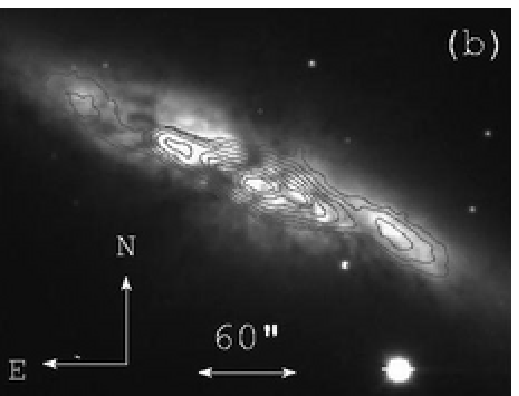}
\caption[]{
(a) A gray-scale representation of the NIR residual image, obtained by 
subtracting an exponential disk from the observed $JHK$ images. 
The central bright 60\arcsec\ corresponds to the bar, with the outer
structures suggesting the presence of a two-armed spiral mode.
(b) The contours of the $K$-band residual image are superposed on a $B$-band 
image. Note that most of the $B$-band knots lie on the spiral arms. A 
prominent dust lane runs along the inner arc of the south-western arm.
The orientation and the image scale are indicated.
See the electronic version for a color version of (a), where
red, green and blue represent $K,H$ and $J$-band residual images, respectively.
}
\end{figure}

\subsection{The azimuthal structure and the spiral arms}

An image of the M\,82 disk was constructed from the exponential scalelength 
and the ellipse-fitting parameters. This was subtracted from the observed 
$JHK$-band images in order to obtain three residual images.  Non-axisymmetric 
structures such as a bar and spiral arms are expected to stand out on these images.
A combined image of these residual images is displayed in Figure~1a.
The most prominent component in this image is a bright linear structure of 
120\arcsec\ length centered around the nucleus, which then extends as arcs out
to a radius of $\sim$160\arcsec. The bar discovered by \citet{Tele91} occupies 
only the central 60\arcsec. The rest of the linear structure, along with the 
arcs represent spiral arm structure viewed almost edge-on, as we demonstrate 
below. The $K$-band residual image contours are superposed on a $B$-band 
image in Figure~1b. Note that the spiral arms are not seen directly 
in $B$-band. However, all the bright blue knots lie along our spiral arms.
\begin{figure} 
\epsscale{1.20}
\vspace{-5mm}
\plotone{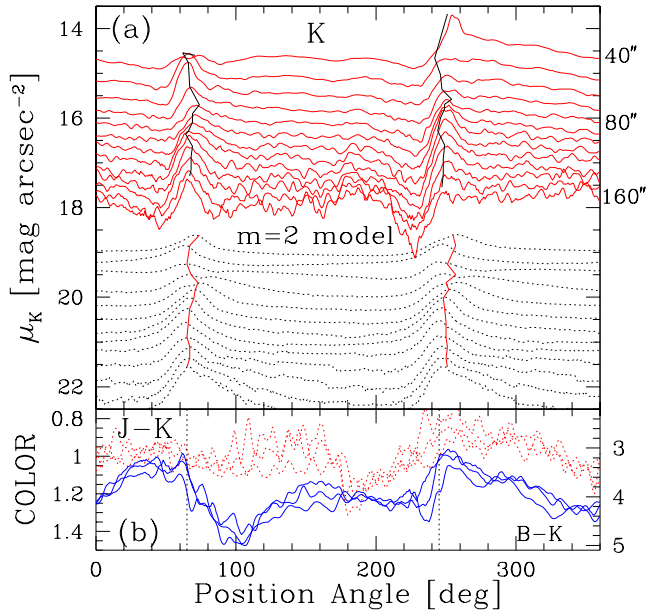}
\caption[]{ 
(a) Azimuthal intensity profiles in the $K$-band image compared to an $m=2$
spiral mode of pitch angle 14$^\circ$ at increasing radial distances between
40\arcsec\ and 160\arcsec\ at intervals of 10\arcsec. The profiles for 
the model are shifted downward by 4 magnitudes for clarity. The spiral 
arms are seen at azimuthal angles $\sim65^\circ$ (north-eastern arm) 
and $\sim245^\circ$ (south-western arm). Almost vertical solid curves at the 
position of the arms on the $K$-band profiles denote the expected 
radius-azimuth dependence for an $m=2$ spiral mode. Similarly the observed 
radius-azimuth relation is superposed on the model profiles.
(b) $J-K$ and $B-K$ (right scale) azimuthal color profiles at three 
representative radii (90\arcsec, 110\arcsec\ and 130\arcsec).
Vertical dotted lines denote the mean position of the spiral arms,
where the observed color gradients are the steepest, the arm colors
being the bluest. See text for details of the spiral model.
}
\end{figure}

A most common technique adopted to characterize the properties of the spiral 
arms is to analyze the azimuthal intensity profiles on deprojected images 
using Fourier analysis. For galaxies with large inclinations, as is the case 
for M\,82, the face-on view of the galaxy cannot be recovered by deprojection. 
Hence, we followed an alternative approach of projecting models of spiral 
galaxies to the inferred inclination angle for M\,82, and comparing the 
azimuthal profiles of these images with those in the $K$-band. 
A face-on model of M\,82 is constructed by adding the light of an $m=2$ 
spiral mode of a given pitch angle to that of an exponential disk. 
Details of this model are given in Section.~4.

The observed azimuthal profiles are compared with model profiles in Figure~2a, 
where azimuthal profiles for radii in the 40\arcsec\ to 160\arcsec\ range
are plotted. The largest plotted radius corresponds to 3.4 times the disk 
scalelength. The north-eastern and south-western arms are easily noticeable
at P.A.s 65$^{\circ}$ and 245$^{\circ}$, respectively. As expected
for a logarithmic spiral, the azimuthal angles of the arms vary systematically 
with radius. In our case, the azimuthal angle turns over at $\approx$80\arcsec\ 
radius. The plotted profiles for the $m=2$ model illustrate that this dependence 
can be traced even in highly inclined systems. The observed azimuthal profiles 
at various radii are very well reproduced by a two-armed spiral mode with 
a 14$^{\circ}$ pitch angle.

The azimuthal profiles are also useful in quantifying the spiral arm contrasts. 
In the $K$-band, the arms are brighter by about $0.50\pm0.1$ magnitudes 
with respect to the inter-arm regions. Corresponding value for the $B$-band
is $\approx1.00\pm0.3$\,mag.  These values translate to contrast parameters
$A_K=1.6\pm0.15$ and $A_B=2.5\pm0.7$, following the \citet{Elme84} definition,
i.e. $A_\lambda=10^{0.4\Delta m_\lambda}$. These quantities in M\,82 are
a factor of four lower than those observed for the grand design spiral M\,51.

The azimuthal profiles of $J-K$ and $B-K$ colors are plotted in Figure~2b. 
The arms show up as the bluest regions, especially in the $B-K$ color. The 
south-western arm is bluer by 1.0~mag in $B-K$ and 0.2~mag in $J-K$ color. 
We also find that the location of the largest disk structures detected in 
near ultraviolet (2500\,\AA) images \citep{Marc01}, coincide with the 
location of the spiral arms. On the left side of the south-western arm, 
the $B-K$ color increases very sharply. This is due to the dust lane present 
on the concave side of the arm (seen also in Figure~1b). The azimuth range 
65--245 corresponds to the dustier part of the galaxy, also affected by the 
superwind cone. 

\section{The nature and origin of the spiral arms}

Why were the spiral arms in M\,82 overlooked in previous studies in
spite of the galaxy's high surface brightness and proximity?
The factors that play against an easy detection are the (i) high inclination 
angle of the disk, (ii) large optical obscuration, and (iii) low contrast
for the spiral arms. This combination prevents the detection of the spiral
structures even after the advent of NIR detectors. We now investigate 
the possible nature of the newly discovered arms. Are they trailing in nature 
as in the majority of galaxies (e.g., de Vaucouleurs, 1958)? 
Do they emerge from the tips of the bar?

In order to answer these questions, we constructed a model galaxy by adding
the light of all relevant sub-components known to exist in M\,82.
The model spiral galaxy contains three main components: an exponential disk, 
an $m=2$ mode spiral and a bar. The 60\arcsec-length bar is tilted with 
respect to the galaxy major axis by 4$^\circ$ \citep{Tele91}. We have adopted 
such a configuration in our model. A fourth component, a starburst nucleus,
modeled as a de Vaucouleurs' profile of effective radius 15\arcsec, was also 
added. The assumed intensity profiles of the bar and the starburst nucleus, 
do not affect our azimuthal profiles, since we limit our analysis to radii 
larger than the bar length. However, the inclusion of the bar in our model 
helps us to investigate the question of whether or not the spiral 
arms emanate from their tips. The spiral arms are chosen to be logarithmic in 
nature, and are generated for radii $>$30\arcsec. The intensity along the arms 
follows the same exponential law of the disk and hence the contrast of the 
arms remains constant all along them. Different model galaxies were generated 
by varying the pitch angle, the initial phase, and the contrast of the spiral 
arms. The parameters of the model that best reproduce the observed $K$-band 
azimuthal intensity profile (i.e. both the observed radius-azimuth and 
radius-contrast relation for the arms) are chosen as the final ones.
The plots for the model in Figure~2a correspond to the best-fit case, 
whose parameters are listed in Table~1, along with the observed values and
errors on them. The error assigned to model parameters indicate the 
range over which acceptable fits can be obtained. 
\begin{center}
\begin{deluxetable}{lll}
\tablewidth{0pc}
\tablecaption{Properties of galactic sub-components}
\tablehead{
\colhead{Component} & \colhead{Observed} & \colhead{Derived} \\
}
\startdata
Inclination [$^{\circ}$] & 77$\pm3$ & 76$\pm1$ \\
Major axis P.A. [$^{\circ}$] & 62--65 & 64$\pm1$ \\
Disk scalelength [\arcsec] & 47$\pm3$ &  48\tablenotemark{a} \\
\hspace{2mm} $\mu_K$(0) [mag\,arcsec$^{-2}$]  & 14.1 & 15.5\tablenotemark{a} \\
Bar length  [\arcsec]     & 60       & 60 \\
Spiral ($m=2$)   &  \nodata        & trailing, logarithmic \\
\hspace{2mm}   \nodata       &  \nodata        & pitch angle=14$\pm1^{\circ}$ \\
    \hspace{2mm} Contrast $A_K$ & $1.6\pm0.15$    & 1.20\tablenotemark{a}      \\
\hline
\enddata
\tablenotetext{a}{Face-on values}
\end{deluxetable}
\end{center}

The face-on, and projected views of the best-matched model galaxy, after 
subtracting the exponential disk, are displayed in Figure~3. It can be seen 
that the spiral arms emerge from the ends of the bar\footnote{For the best-fit 
model with constant pitch angle, shown in Figure~3, 
there is a phase difference of $\sim20^{\circ}$ between the tips of the bar 
and the beginning of the arms. However, the arms can be connected to the tips
of the bar by allowing the pitch angle to vary smoothly along the arms from 
13$^{\circ}$ in the inner part to 15$^{\circ}$ at radius=160\arcsec.
},
as is commonly observed in barred galaxies.
Arms wind $\sim360^\circ$ before they can be noticed as such on the residual 
image at $\sim$80\arcsec\ radius. On the projected image, the arms are almost 
straight features in the 30--60\arcsec\ radial zone, appearing as an 
extention of the bar. The spiral arms are not noticeable unless an exponential
disk is subtracted.
\begin{figure}
\epsscale{1.15}
\plotone{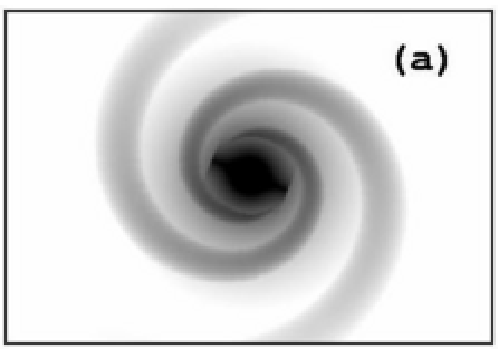}
\plotone{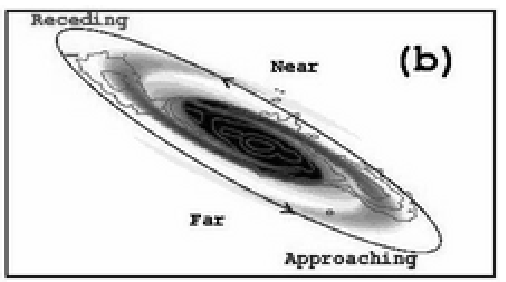}
\caption[]{
The face-on, and projected views of the best-matched model galaxy, after
subtracting the exponential disk. $K$-band residual image contours are 
superposed on the latter image.
The ellipse with arrows indicates the sense of rotation of stars expected
for trailing spiral arms.
}
\end{figure}

The expected sense of rotation of stars in trailing spiral arms is indicated 
in Figure~3. The observed geometrical and kinematical orientations of the
galaxy are also shown. These are based on the scheme adopted by \citet{Tele91}. 
This configuration is consistent with the spiral arms being trailing in nature. 
The presence of a dust lane on the concave side of the south-western arm also 
supports the trailing nature of the arms. Besides, the molecular gas 
distribution in the disk closely follows this dust lane \citep{Walt02}. 
Further out, the molecular gas warps, connecting smoothly with the off-planar 
H\,I streamers reported by \citet{Yun93}. The joint configuration of the bar, 
spiral arms, molecular gas, and H\,I streamers suggest that they are part of 
a global $m=2$ mode excited by a close encounter of M\,82 with its neighbors 
M\,81 and NGC\,3077. The bi-symmetric structure might be facilitating the 
infall of gas from the debris surrounding M\,82 all the way to the nuclear
region, thus fueling the starburst. In summary, M\,82 seems to fit well into 
the picture of spiral formation presented by \citet{Toom72}. 

\section{Concluding Remarks}

The discovery of spiral arms in M\,82 throws new light on the inhomogeneous
class of galaxies known as Irr\,II galaxies. These are disk galaxies that 
differ from spirals in not showing an obvious spiral structure. They differ
from irregulars (Im, Sm) by showing a much redder and unresolved disk, and 
from lenticulars by having an early-type (i.e. younger) stellar spectrum. 
It is well known that starburst activity, dust obscuration and tidal 
interaction play varying degrees of importance in shaping different galaxies 
in this class \citep{Krie74}. All the three phenomena are present in M\,82. 
In fact, the tidal interaction is responsible for both the chaotic 
distribution of dust and triggering the starburst activity. This suggests 
that tidal interaction may play an important role in other galaxies of this 
class as well. \citet{OCon78} have argued that the gross properties of
M\,82 --- its mass, luminosity, size, mean spectral type --- are comparable
to those of normal late type (Sc/Irr) galaxies. The rather small bulge 
(mass$\sim3\times10^7$\msun; Gaffney, Lester \& Telesco 1993), and the relatively open 
arms inferred in the present work (pitch angle of 14$^\circ$), suggest a 
late morphological type SBc for M\,82.

\vspace*{-0.6cm}

\acknowledgments
It is a pleasure to thank A. Bressan, E. Recillas, O. L\'opez-Cruz,
and I. Puerari for discussions during different phases of this work.
The presentation of the paper has vastly improved following the insights
given by the referee. We thank G. Escobedo and all the staff at OAGH, 
especially E. Castillo, for technical help during observations. 
This work is partly supported by CONACyT (Mexico) research grants
G28586-E and 39714-F. This research has made use of the NED, which is 
operated by the JPL/NASA.


\end{document}